\documentclass[]{spie}  %>>> use for US letter paper
\usepackage{amsmath,amsfonts,amssymb}
\usepackage{graphicx}
\usepackage[colorlinks=true, allcolors=blue]{hyperref}
\usepackage{bm}

\title{Tip-tilt anisoplanatism in MCAO-assisted astrometric observations}

\author[a]{Giulia Carlà}
\author[a]{Lorenzo Busoni}
\author[a]{C{\'e}dric Plantet}
\author[a]{Guido Agapito}
\author[b]{Carmelo Arcidiacono}
\author[c]{Paolo Ciliegi}
\affil[a]{INAF, Osservatorio Astrofisico di Arcetri, Largo Enrico Fermi 5, 50125 Firenze, Italy}
\affil[b]{INAF Osservatorio Astronomico di Padova, Vicolo Osservatorio 5, 35122 Padova, Italy}
\affil[c]{INAF, Osservatorio di Astrofisica e Scienza dello Spazio, Via Gobetti 93/3, 40129 Bologna, Italy}

\authorinfo{Further author information: (Send correspondence to G. C.)\\G. C.: E-mail: giulia.carla@inaf.it, Telephone: +39 055 2752289}
\begin{document} 
\maketitle

\begin{abstract}
A new era of ground-based observations, either in the infrared with the next-generation of 25-40m extremely large telescopes or in the visible with the 8m Very Large Telescope, is going to be assisted by multi-conjugate adaptive optics (MCAO) to restore the unprecedented resolutions potentially available for these systems in absence of atmospheric turbulence. Astrometry is one of the main science drivers, as MCAO can provide good quality and uniform correction over wide field of views ($\sim$ 1 arcmin) and offer a large number of reference sources with high image quality. The requirements have been set to very high precisions on the differential astrometry (e.g. 50µas for MICADO/MORFEO - formerly known as MAORY - at the Extremely Large Telescope) and an accurate analysis of the astrometric error budget is needed. In this context, we present an analysis of the impact of MCAO atmospheric tip-tilt residuals on relative astrometry. We focus on the effects of the scientific integration time on tip-tilt residuals, that we model through the temporal transfer function of the exposure. We define intra- and inter-exposure tip-tilt residuals that we use in the estimation of the centroiding error and the differential tilt jitter error within the astrometric error budget. As a case study, we apply our results in the context of the MORFEO astrometric error budget.
\end{abstract}


\section{INTRODUCTION}
\label{sec:intro}
The next generation of ground-based observations with the extremely large telescopes \cite{Tamai20,Fanson20,Sanders13} foresees the use of adaptive optics (AO) systems to compensate for the wavefront distortions caused by the atmospheric turbulence. Multi-conjugate adaptive optics (MCAO) \cite{Rigaut18} will be a key element to provide images close to the diffraction limit over wide fields of view ($\sim$1arcmin), thanks to use of multiple guide stars and deformable mirrors (DMs) to perform a tomographic reconstruction of the turbulent volume and compensate for different layers of the atmosphere. Relative astrometry is one of the main science drivers of the instruments equipped with MCAO, as it can benefit from the uniform correction, the large number of reference sources with high image quality and the control of plate-scale distortions that this flavour of AO is able to provide. These characteristics, together with the high resolution of the future observations, motivate the challenging requirements that have been set for the precision on the differential astrometry. The Multiconjugate adaptive Optic Relay For ELT Observations (MORFEO, formerly known as MAORY)\cite{Ciliegi22}, that will equip the Multi-AO Imaging CamerA for Deep Observations (MICADO)\cite{Davies21} at the Extremely Large Telescope (ELT), is required to deliver precisions of 50µas with a goal of 10µas\cite{Rodeghiero19}. In this context, an accurate analysis of the astrometric error budget is needed in order to keep the errors under control during both the observations preparation and the post-processing phase. Among the sources of astrometric error that can affect MCAO-assisted observations \cite{Trippe10}, we focus our analysis on the errors that can be influenced by tip-tilt residuals, in particular the differential tilt jitter error \cite{Trippe10,Cameron09,Fritz10} and the centroiding error \cite{Lindegren78}. In Ref.~\citenum{Carla22}, we presented an analytical formalism to compute the residual phase from an MCAO loop and we derived an expression to estimate the temporal power spectrum of the residuals in any direction of the scientific field of view. The formulation has been used for a first characterization of the behavior of MCAO tip-tilt residuals and the related impact on astrometry. In this work, we develop the analysis on the temporal effects of the scientific integration on tip-tilt residuals. We identify intra- and inter-exposure tip-tilt residuals, that we relate to both centroiding and differential tilt jitter error. We then use our results to investigate the contribution of atmospheric tip-tilt residuals to the MORFEO astrometric error budget.\newline
In Sec.~\ref{sec:temp_behavior_tt_res}, we define the intra- and inter-exposure tip-tilt residuals and we model them in the temporal frequency domain; in Sec.~\ref{sec:tt_astrometry}, we use intra- and inter-exposure tip-tilt residuals to estimate differential tilt jitter and centroiding error; in Sec.~\ref{sec:mao_tt_res}, we estimate the contribution of atmospheric tip-tilt residuals to the MORFEO astrometric error budget.

\section{Temporal behavior of tip-tilt residuals}
\begin{figure}[htbp]
    \centering
    \includegraphics[width=.9\linewidth]{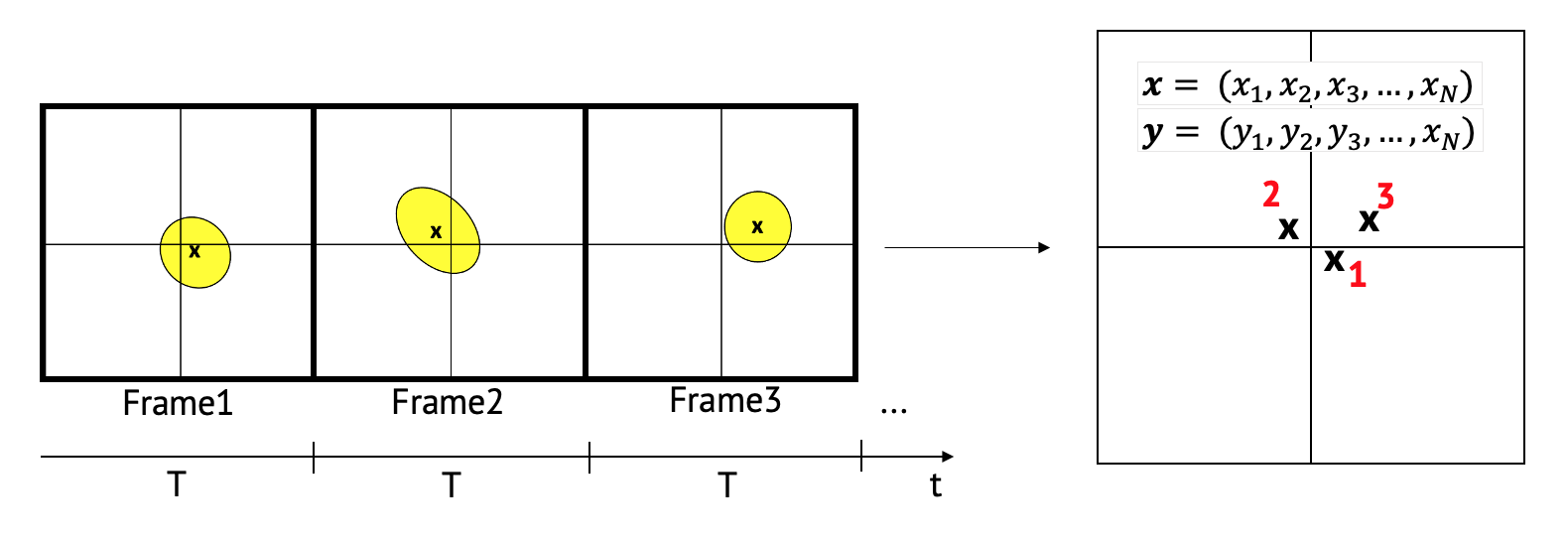}
    \caption{Schematic representation of the effects of tip-tilt residuals on exposures of integration time $T$. The fluctuations integrated during an exposure can result in an elongation of the PSF (yellow spots); the fluctuations that are not integrated can determine a variation of the PSF position between successive frames (black crosses).}
    \label{fig:temp_res}
\end{figure}

\begin{figure}[htbp]
    \centering
    \includegraphics[width=0.95\linewidth]{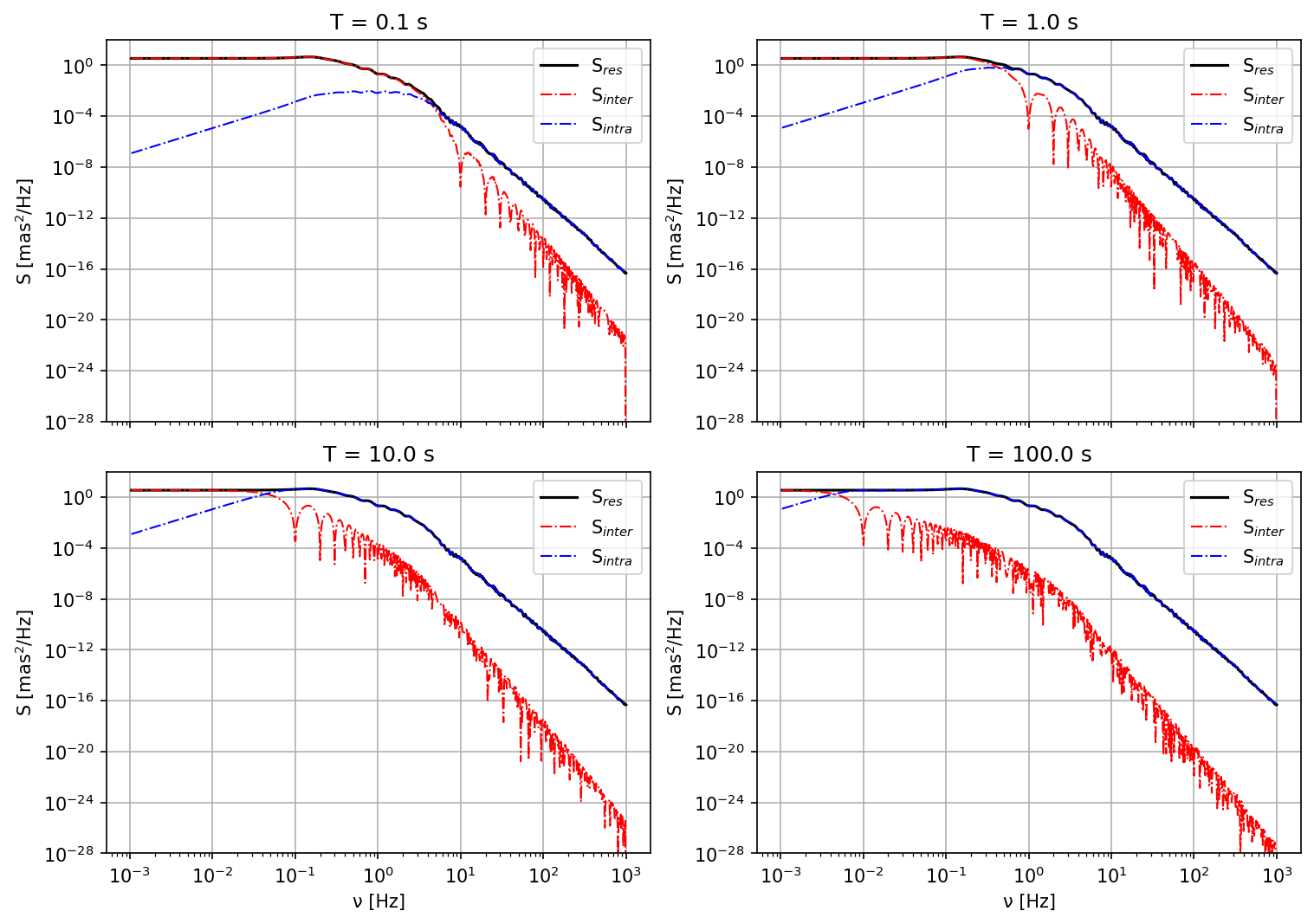}
    \caption{Temporal PSD of total (black line), inter- (red line) and intra-exposure (blue line) tip-tilt residuals as a function of the temporal frequencies. The exposure time ranges from 0.1 (top left) to 100s (bottom right). The PSDs are computed for a source on axis, considering a telescope aperture of 39m, a natural guide stars asterism radius of 80", a zenith angle of 30$^{\circ}$, two DMs conjugated at 600m and 17km and the ELT median turbulence profile reported in Ref.~\citenum{elt_profile}.}
    \label{fig:inter_intra_res_psds}
\end{figure}

\begin{figure}[ht]
    \centering
    \includegraphics[width=\linewidth]{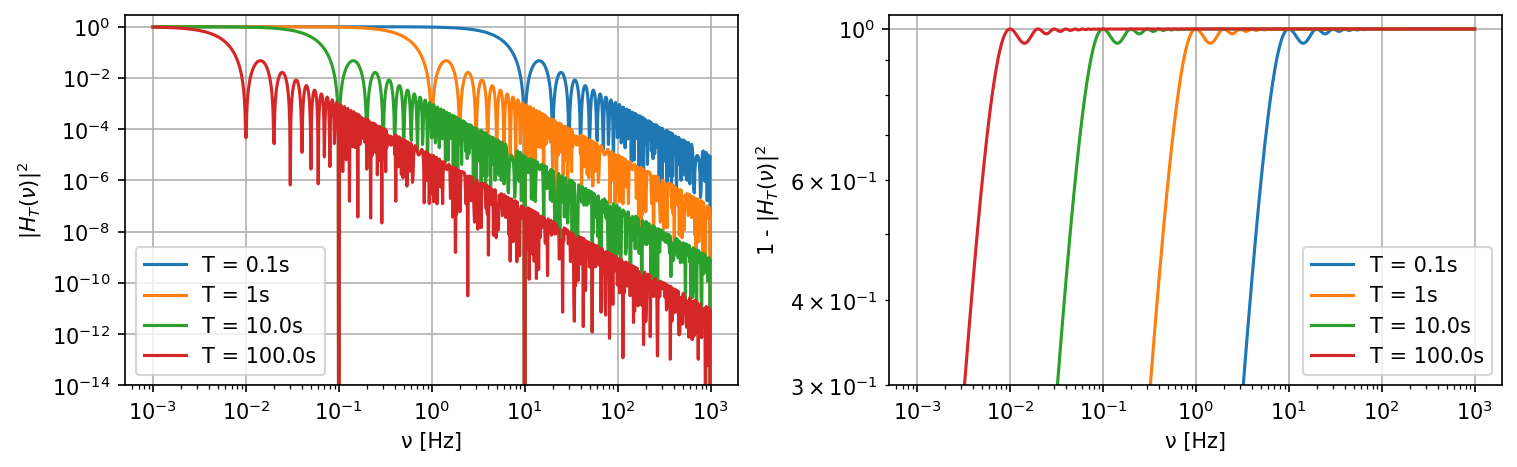}
    \caption{Temporal transfer function of the inter- (left) and intra-exposure (right) tip-tilt residuals as a function of the temporal frequencies. The colors represent different exposure times in a range from 0.1 to 100s. The cut-off frequency $\nu_{cutoff}$ = 1/T determines when either the inter-exposure transfer function starts filtering, or the intra-exposure transfer function stops filtering the temporal PSD of tip-tilt residuals.}
    \label{fig:camera_tfs}
\end{figure}

\begin{figure}[ht]
    \centering
    \includegraphics[width=0.6\linewidth]{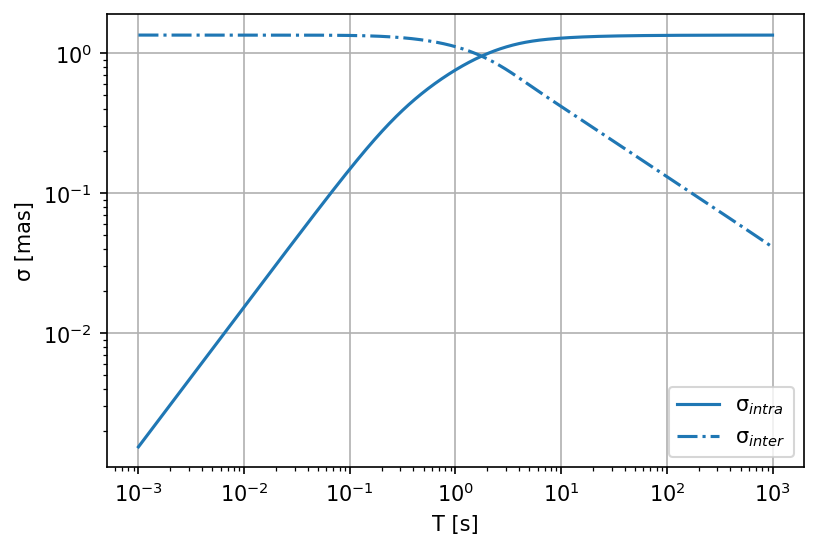}
    \caption{Intra- (solid line) and inter-exposure (dash-dotted line) tip-tilt error as a function of the exposure time. The curves are obtained in the same configuration as Fig.~\ref{fig:inter_intra_res_psds}.}
    \label{fig:inter_intra_res_std}
\end{figure}

% \section{Temporal spectrum of intra- and inter-exposure tip-tilt residuals}
% \section{Intra- and inter-exposure tip-tilt residuals}
\label{sec:temp_behavior_tt_res}
Tip-tilt residuals introduce uncertainties in the astrometric measurements, as they determine fluctuations of the position of a source with respect to its nominal position on the detector. They can lead to two effects: on the one hand, the residuals integrated during a single exposure can result in an elongation of the point spread function (PSF) observed on the image; on the other hand, the amount of fluctuations that is not integrated within the exposure can be observed as a jitter of the source position between successive frames. We identify the former residuals as \textit{intra-exposure} and the latter as \textit{inter-exposure} (Fig.~\ref{fig:temp_res}).
The sum of intra- and inter-exposure residuals spectra gives the total residual spectrum of tip-tilt:
\begin{equation}
    \label{eq:tot_res}
    S_{res}^{\alpha}(\nu) = S_{intra}^{\alpha}(\nu) + S_{inter}^{\alpha}(\nu) \, ,
\end{equation}
where $S$ denotes the temporal PSD of tip-tilt and $\alpha$ indicates a specific direction of the field of view. An expression to compute $S_{res}^{\alpha}$ in the case of an MCAO loop has been derived in Ref.~\citenum{Carla22}. The temporal power spectrum of the inter-exposure residuals can be obtained from the product of the temporal PSD of the residuals and the square modulus of the transfer function $H_T$ representing the scientific integration over a temporal length $T$:
\begin{equation}
\begin{split}
\label{eq:inter_exp_res}
    S_{inter}^{\alpha}(\nu) &= |H_{T}(\nu)|^2 S_{res}^{\alpha}(\nu) \\
    &= H_{T, inter}(\nu) S_{res}^{\alpha}(\nu) \, ,
\end{split}
\end{equation}
where we have identified $H_{T, inter}$ as the temporal transfer function of the inter-exposure residuals.
From Eq.~\eqref{eq:tot_res} and Eq.~\eqref{eq:inter_exp_res}, we derive an expression to estimate the temporal PSD of the intra-exposure residuals temporal spectrum as well:
\begin{equation}
\begin{split}
\label{eq:intra_exp_res}
    S_{intra}^{\alpha}(\nu) &= \big(1 - |H_{T}(\nu)|^2 \big) S_{res}^{\alpha}(\nu) \\
    &= H_{T, intra}(\nu) S_{res}^{\alpha}(\nu) \, ,
\end{split}
\end{equation}
where we have introduced $H_{T, intra}$ as the temporal transfer function of the intra-exposure residuals.\newline
In Fig.~\ref{fig:inter_intra_res_psds}, we plot the temporal PSD of both total, inter-exposure and intra-exposure tip-tilt residuals. The exposure time ranges from 0.1 to 100s.
The curves show that the inter-exposure transfer function filters out the high temporal frequencies that are, on the other hand, passed by the intra-exposure transfer function. The amount of temporal frequencies that each transfer function filters out depends on the integration time of the scientific exposure that determines the transfer functions cut-off frequency ($\nu_{cutoff}$ = 1/T), as we can see in Fig.~\ref{fig:camera_tfs}.
This behavior leads to the dependence of intra- and inter-exposure tip-tilt residuals on the integration time that is shown in Fig.~\ref{fig:inter_intra_res_std}.
Increasing the integration time helps to reduce the residual jitter that is observed between successive frames ($T^{-1/2}$ dependence for times larger than $\sim$1s) but, on the other hand, determines larger intra-exposure residuals ($T^1$ dependence for times smaller than $\sim$1s) that can affect the PSF shape and size. In the next section, we analyze how these effects take part in the error budget of relative astrometry.

\section{Impact of MCAO tip-tilt residuals on relative astrometry}
\label{sec:tt_astrometry}
Let us consider the measurement of the distance between two objects at positions $\alpha$ and $\beta$. Tip-tilt residuals can affect the precision on the relative astrometric measurements through two effects: differential tilt jitter and centroiding error. Differential tilt jitter represents fluctuations of the distance measurements observed between successive frames and is caused by the difference between inter-exposure tip-tilt residuals in the two considered directions \cite{Carla22}. The centroiding error represents the theoretical limit to the astrometric precision that is due to the photon noise and depends on both the dimension of the PSF and the signal-to-noise ratio (SNR) \cite{Lindegren78}:
\begin{equation}
    \label{eq:lindegren_eq}
    \sigma_{cent} \sim \dfrac{1}{\pi}\dfrac{FWHM}{SNR} \, ,
\end{equation}
where FWHM is the full width at half maximum, that can be affected by the intra-exposure tip-tilt residuals. Considering these effects, the uncertainty on the relative astrometric measurements due to tip-tilt residuals can be estimated as\cite{Cameron09}:
\begin{equation}
    \label{eq:astrom_error}
    (\sigma_d^{\alpha,\beta})^2 = (\sigma_{DTJ}^{\alpha,\beta})^2 + (\sigma_{cent}^{\alpha})^2 + (\sigma_{cent}^{\beta})^2 \, ,
\end{equation}
where $(\sigma_d^{\alpha,\beta})^2$ is the variance of the distance measurements between $\alpha$ and $\beta$, $(\sigma_{DTJ}^{\alpha,\beta})^2$ is the variance of differential tilt jitter and $(\sigma_{cent}^x)^2$ indicates the variance due to the centroiding error in a direction $x$ ($x$ = $\alpha$, $\beta$) of the field of view. In the following, we estimate the contribution of both differential tilt jitter and centroiding error in a MORFEO-like configuration: we consider a telescope aperture of 39m, two DMs conjugated at 600m and 17km, an equilateral asterism of natural guide stars centered at the origin of the field of view and the ELT median turbulence profile reported in Ref.~\citenum{elt_profile}, with a zenith angle of 30$^{\circ}$. We consider a loop with a frequency frame rate of 1kHz, where the control is a pure integrator. We compute the tomographic reconstructor through a simple Least Square Error (LSE) estimator\cite{Madec99,Carla22}, starting from the tip-tilt measurements on the three natural guide stars wavefront sensors, to compensate for tip-tilt on the first DM and focus-astigmatisms on the second DM. In this section, we do not consider the noise on the wavefront sensors measurements.

\subsection{Differential tilt jitter error}
\label{sec:dtj}
\begin{figure}[ht]
\centering
\begin{minipage}{.49\textwidth}
  \centering
  \includegraphics[width=\linewidth]{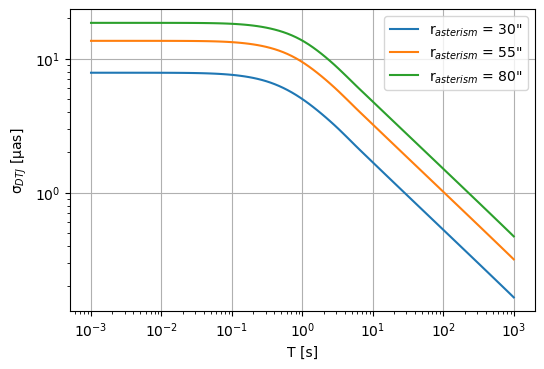}
  \caption{Differential tilt jitter error as a function of the exposure time for two objects separated by 1". The curves are obtained in the case of a MORFEO-like configuration as described in the text. The colors represent different values of the asterism radius.}
  \label{fig:mao_dtj_inter_exp_res}
\end{minipage}\hfill
% \hspace{.1cm}
\begin{minipage}{.49\textwidth}
  \centering
  \includegraphics[width=\linewidth]{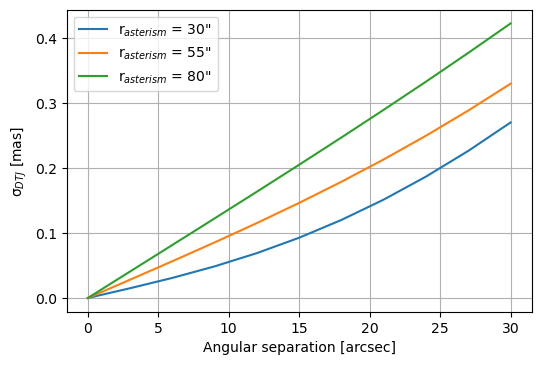}
  \caption{Differential tilt jitter error as a function of the angular separation of the sources, for an exposure time of 1s. The curves are obtained in the case of a MORFEO-like configuration as described in the text. The colors represent different values of the asterism radius.}
  \label{fig:mao_dtj_vs_ast_dist}
\end{minipage}
\end{figure}

The amount of differential residual jitter that is left between successive frames can be described as the inter-exposure residuals and can be modeled through Eq.~\eqref{eq:inter_exp_res} by substituting $S_{res}^{\alpha}$ with the temporal PSD of differential tilt jitter. The variance can then be obtained from the integration of the PSD over the temporal frequencies. We use the expression of the differential tilt jitter temporal PSD derived in Ref.~\citenum{Carla22} (Eqs.~(29)-(31)) in the case of an MCAO correction to estimate the differential tilt jitter error in the MCAO configuration described above. In Fig.~\ref{fig:mao_dtj_inter_exp_res}, we show the differential tilt jitter error as a function of the exposure time, for different values of the asterism radius. 
The dependence on the exposure time as $T^{-1/2}$ (for times larger than $\sim$1s in this configuration) shows that the astrometric error due to differential tilt jitter can be controlled through proper integration times. In Fig.~\ref{fig:mao_dtj_vs_ast_dist}, we show the differential tilt jitter error as a function of the distance between the two astrometric sources, for an exposure time of 1s. The dependence of MCAO differential tilt jitter error on the distance depends on the position of the sources with respect to the asterism of guide stars and it becomes approximately linear when the asterism radius is significantly larger than the scientific field.

\subsection{Centroiding error}
\begin{figure}[ht]
    \centering
    \includegraphics[width=0.6\linewidth]{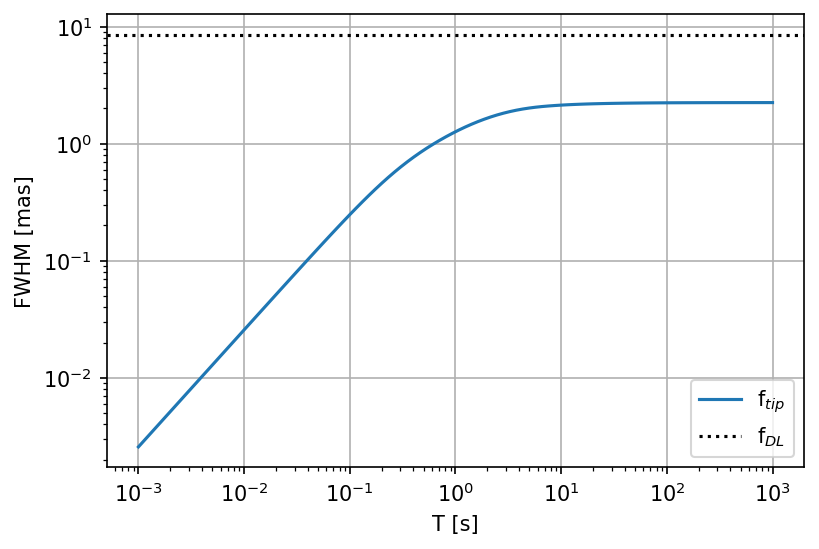}
    \caption{FWHM of the gaussian kernel related to intra-exposure residuals of tip (solid line), for a source on axis, as a function of the exposure time. The value of the diffraction-limited FWHM in the H-band is shown in comparison (dotted line). The configuration is the same as Fig.~\ref{fig:mao_dtj_inter_exp_res}, with an asterism radius of 80".}
    \label{fig:mao_fwhm}
\end{figure}

\begin{figure}[ht]
    \centering
    \includegraphics[width=0.6\linewidth]{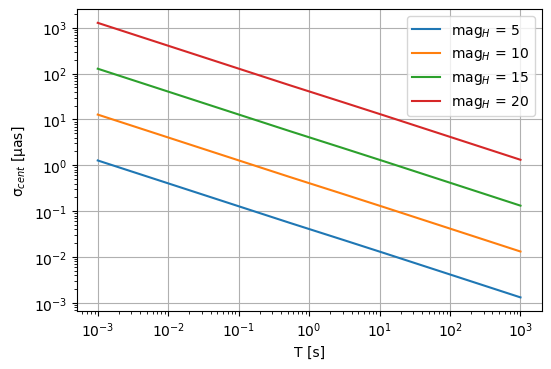}
    \caption{Centroiding error as a function of the exposure time, computed for an on-axis target. The colors show different values of the magnitude of the source in the H-band. The configuration is the same as Fig.~\ref{fig:mao_dtj_inter_exp_res}, with an asterism radius of 80".}
    \label{fig:mao_cent_error}
\end{figure}

The intra-exposure tip-tilt residuals can affect the centroiding error as they can impact on the PSF shape and size. In this case, the effect has to be taken into account within the FWHM computation. Assuming tip-tilt residuals approximated through Gaussian statistics \cite{Olivier93}, we can estimate the PSF from the convolution between the diffraction-limited PSF and the gaussian kernel due to tip-tilt residuals. We can then derive the FWHM as:
\begin{align}
    \label{eq:fwhm}
    f_{conv, x} &= \sqrt{f_{DL}^2 + f_{tip}^2} \\
    f_{conv, y} &= \sqrt{f_{DL}^2 + f_{tilt}^2} \, ,
\end{align}
where $f_{conv}$ is the FWHM of the convolved PSF ($x$ and $y$ indicate the two axes), $f_{DL}$ is the FWHM of the diffraction-limited PSF and $f_{tip (tilt)}$ is the FWHM computed from the standard deviation related to the intra-exposure tip(tilt) residuals.
In Fig.~\ref{fig:mao_fwhm}, we show the FWHM resulting from intra-exposure atmospheric tip-tilt residuals as a function of the integration time for an on-axis source with the MORFEO-like configuration presented in Sec.~\ref{sec:tt_astrometry} and with an NGS asterism radius of 80". The diffraction-limited FWHM for infrared observations at 1.6 $\mu$m (H-band) is shown in comparison. As in Fig.~\ref{fig:inter_intra_res_std}, the plots show that larger integration times lead to a larger impact of the intra-exposure tip-tilt residuals on the FWHM.
However, in this configuration the contribution is smaller than the value of the diffraction-limited FWHM ($\sim$ 8.5mas), which therefore dominates the contribution to the centroiding error. This is evident in Fig.~\ref{fig:mao_cent_error} where the centroiding error, derived from Eq.~\eqref{eq:lindegren_eq} and Eq.~\eqref{eq:fwhm}, is plotted as a function of the exposure time. The behavior is dominated by the $T^{-1/2}$ dependence of the centroiding error given by the diffraction-limited FWHM. Different magnitudes in the H-band have been taken into account, considering a total transmission of 65\%. 

\section{TIP-TILT RESIDUALS IN THE MORFEO ASTROMETRIC ERROR BUDGET}
\begin{figure}[ht]
    \centering
    \includegraphics[width=0.5\linewidth]{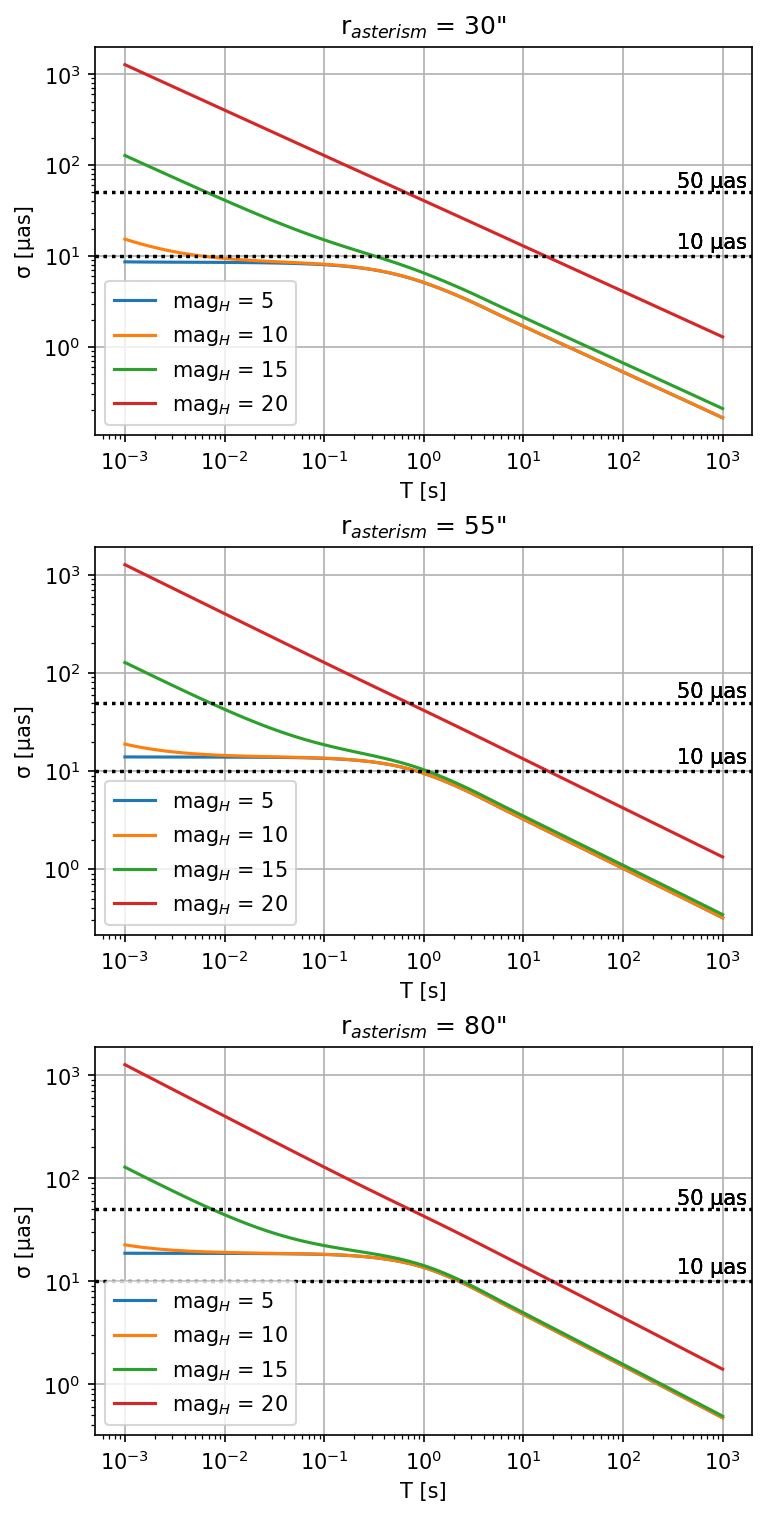}
    \caption{Astrometric error due to atmospheric tip-tilt residuals of MORFEO as a function of the scientific integration time. The errors are shown for an asterism radius of 30" (top), 55" (center) and 80" (bottom); different values of noise have been considered depending on the asterism, as shown in Table~\ref{table:mao_noise_params}. The colors represent different magnitudes of the astrometric sources in the H-band. Both the requirement (50µas) and the goal (10µas) of MORFEO astrometric precision are shown in comparison (dotted black lines).}
    \label{fig:mao_astrom_error}
\end{figure}

We estimate the contribution of atmospheric tip-tilt residuals to the MORFEO astrometric error budget, by means of Eq.~\ref{eq:astrom_error}. We compute the error in the case of two sources with an angular separation of 1", placed at the center of the field of view. We consider an equilateral asterism of natural guide stars, that we assume of $mag_H$=18, and we now consider the noise on the wavefront sensors measurements. The related parameters, that are based on the sky coverage assessment for MORFEO \cite{Plantet19}, are shown in Table~\ref{table:mao_noise_params}. The configuration is the same as the one considered in the previous section for what concerns the telescope aperture, the conjugation height of the DMs and the turbulence profile. In Fig.~\ref{fig:mao_astrom_error}, we show the astrometric error as a function of the exposure time, for different values of the astrometric targets magnitude in the H-band. The plots show that the centroiding error dominates the contribution to the astrometric error for the fainter sources. On the other hand, the high SNR of the brighter objects ($mag_H$ = 5, 10) makes the centroiding error negligible with respect to the differential tilt jitter effect. However, in both cases, the indication is to integrate in order to keep the errors under control, since they can be reduced with the exposure time as $T^{-1/2}$ (for times larger than $\sim$ 1s in case the differential tilt jitter error is dominant). The two considered sources of astrometric error should not represent a relevant contribution to the MORFEO astrometric error budget for sources at an angular separation of 1". Though, these have to be considered preliminary results: in our estimation we have taken into account noise, temporal and tomographic errors within the MCAO loop, but we have neglected other contributors such as aliasing and high-order residuals.
\vspace{.5cm}
\begin{table}[ht]
% \small
\centering
\renewcommand{\arraystretch}{1.3}
\begin{tabular}{ |c|c c c| } 
 \hline
 $r_{asterism}$ ["] & 30 & 55 & 80 \\
 \hline
 $\sigma_{noise}$ [nm] & 100 & 164 & 250 \\ 
 \hline
 $\nu_{loop}$ [Hz] & 500 & 500 & 500 \\
   \hline
 $g$ & 0.1 & 0.1 & 0.15 \\
 \hline
 $d$ & 3 & 3 & 3 \\
 \hline
\end{tabular}
\vspace{.5cm}
\caption{Parameters used for the MCAO loop. $\nu_{loop}$ is the loop frequency frame rate, $g$ is the loop gain and $d$ is the total delay in frames.}
\label{table:mao_noise_params}
\end{table}

\label{sec:mao_tt_res}

\section{Conclusions}
We have presented an analysis on the effects of atmospheric tip-tilt residuals on astrometric observations with ground-based telescopes equipped with MCAO. We have modeled the impact of the scientific exposure on tip-tilt residuals that we have described through the definition of proper temporal transfer functions. We have identified intra- and inter-exposure tip-tilt residuals and we have shown the behavior of the former as high-pass filter and of the latter as low-pass filter of the temporal spectrum of the tip-tilt residuals. We have related the two effects, respectively, with the variation of the PSF shape within a single exposure and the jitter of the PSF centroid between successive exposures. We have used our results to model two sources of astrometric error, the centroiding error and the differential tilt jitter error. Despite the linear dependence of intra-exposure residuals on the scientific exposure time, in the considered configuration the centroiding error shows a dependence on the inverse of the square root of the scientific exposure time, as it is dominated by the contribution of the diffraction-limited FWHM. The same dependence on the integration time characterizes the differential tilt jitter error, for times larger than $\sim$1s in the considered configuration. We have presented an estimation of the contribution of atmospheric tip-tilt residuals to the astrometric error budget of MORFEO and we have shown that, for both the centroiding and the differential tilt jitter error, the integration time can be a key aspect to keep the errors under control. For proper integration times, atmospheric tip-tilt residuals should not be relevant within the MORFEO astrometric error budget, in the case of sources separated by 1".

% References
\bibliography{spie2022} % bibliography data in report.bib
\bibliographystyle{spiebib} % makes bibtex use spiebib.bst

\end{document}